\documentclass[amsmath,amssymb,superscriptaddress,aps,pre,reprint]{revtex4-2}

\usepackage[T1]{fontenc}
\usepackage{lmodern}
\usepackage[english]{babel}
\usepackage[utf8]{inputenc}
\usepackage{bm}
\usepackage{amsmath}
\usepackage{xcolor}
\usepackage{graphicx}

\newcommand{\av}[1]{\left< #1 \right>}
\newcommand{\kbT}{k_\text{B}T}
\newcommand{\Fex}{\bm{F}_\text{ex}}

\newcommand{\Fi}{\bm{F}_i}

\renewcommand{\vr}{\bm{r}}
\newcommand{\vk}{\bm{k}}
\newcommand{\vq}{\bm{q}}
\newcommand{\vp}{\bm{p}}
\newcommand{\eq}{\text{eq}}

\newcommand{\ex}{\text{ex}}
\newcommand{\irr}{\text{irr}}

\newcommand{\del}{\bm{\partial}}
\newcommand{\proq}{\mathcal{Q}}
\newcommand{\prop}{\mathcal{P}}

\newcommand{\Delt}{\Delta t}

\definecolor{ao}{rgb}{0, 0.5, 0}

\begin{document}

\title{Nonlinear microrheology with time-dependent forces - Application to recoils in viscoelastic fluids}
\date{\today}
\author{Nikolas Ditz}
\affiliation{Fachbereich Physik, Universit\"at Konstanz, 78457 Konstanz, Germany}
\author{Antonio M. Puertas}
\affiliation{Departamento de Qu\'{\i}mica y F\'{\i}sica, Universidad de Almer\'{\i}a, 04.120 Almer\'{\i}a, Spain}
\author{Matthias Fuchs}
\affiliation{Fachbereich Physik, Universit\"at Konstanz, 78457 Konstanz, Germany}

\begin{abstract}
This work presents a theoretical analysis of the motion of a tracer colloid driven by a time-dependent force through  a viscoelastic fluid. The recoil of the colloid after application of a strong force is determined. It provides insights into the elastic forces stored locally in the fluid and their weakening by plastic processes. We generalize  the mode coupling theory of microrheology  to include time-dependent forces. After deriving the equations of motion for the tracer correlator and simplifying to a schematic model we apply the theory to a switch-off force protocol that features the recoiling of the tracer after cessation of the driving. We also include Langevin dynamics simulations to compare to the results of the theory.  A non-monotonic trend of the recoil amplitude is found in the theory and confirmed in the simulations. The linear-response approximation is also verified in the small-force regime.
While the overall agreement between simulation and theory is good, simulation shows that the theory predicts a too strong non-monotonous dependence of the recoil distance on the applied force.

\end{abstract} 

\maketitle


\section{Introduction}

Complex fluids are known to possess viscous and elastic materials properties. They are able to flow when stirred slowly and to respond elastically when tested at high frequencies. Viscoelasticity on the macroscopic length scale has amply been observed and discussed e.g.  in rheological experiments and theory \cite{Larson,Wagner2021}. On the microscopic scale, observing the motion of a tracer particle embedded in the complex fluid has also revealed important insights. On the one hand, the jittery motion of the fluctuating tracer can be considered as a stochastic process in the heat bath provided by the fluid. Classic stochastic models, like the Brownian random walk, have been introduced and developed in this setting \cite{Dhont,vanKampen}. On the other hand, the tracer  can be used as a probe particle for studying the complex fluid itself \cite{FurstBook, Puertas2014}. For both cases, often colloids are used as probe particles for testing viscoelastic phenomena in complex fluids including in biological samples. This can be done passively, using thermally fluctuating tracers, or actively by forcing the colloidal probe through the medium. Different modes of driving, e.g.~at constant force or constant velocity, are also possible \cite{Squires2005}.

A specific microrheological protocol has been developed recently, that provides unique information on the elastic forces stored in complex fluids 
\cite{Gomez-Solano2015,Ginot2022barrier,Ginot2022recoil,KrishnaKumar2023}. Forcing a colloidal probe and then releasing it, a back-motion of the colloid has been recorded after its release. The recoil results from that part of the force field, which has developed in the fluid during the forced motion of the tracer, that is of elastic nature and  pushes back the tracer   \cite{Gomez-Solano2015,Ginot2022recoil}. This effect has also already been seen in simulations \cite{Zia2013}.  Measuring the recoil motion  gives access to rapid structural process in the viscoelastic fluid, which progress much faster than Maxwell's viscous relaxation. Relating the recoil motion to the familiar mean squared displacement in the linear response regime of equilibrium fluctuations has supported this interpretation. Fast processes, where the particles have moved only little, dominate the recoil motion, while the long-time diffusion does not appear \cite{Caspers2023}. 

The present contribution is aimed at developing a theoretical approach to the recoil motion of a colloidal tracer after driving by a strong force, viz.~the nonlinear recoil, where the recoil motion depends on the strength of the forcing before switch-off. The approach builds on the microscopic mode coupling theory of active microrheology which has captured the nonlinear velocity-force relations and the distributions of tracer displacements in viscoelastic fluid and soft solid states 
\cite{Gazuz2009,Gazuz2013,Harrer2012,Gnann2011,GruberDiss,Gruber2020}. We generalize this approach to general time-dependent forcing, and apply the developed theory to recoil, viz.~a force protocol where a constant force is applied for a finite window in time. Additionally, we perform Langevin dynamics simulations to test the theory and to determine recoil spectra in simulations of dense fluids. The generalization to time-dependent forces transfers techniques developed for macroscopically sheared dispersions \cite{Voigtmann2012,Brader2012}, where e.g.~the stress response after switching-off a strong shearing 
has been studied \cite{PRL2013}.

An interesting class of viscoelastic fluids is given by
dense colloidal dispersions. They can be prepared as well characterized model fluids, e.g.~of hard sphere like particles, and the viscoelastic phenomena can widely be varied by finely tuning   the distance to the colloidal glass transition  \cite{Pusey,Hunter2012}. Their macroscopic rheology has been studied intensely \cite{Wagner2021,Siebenbuerger2012}
The caging of particles in the shell of neighbors which themselves are hindered by the original particle has emerged as a central nonlinear mechanism causing elastic recoil forces and an increase of the dispersion viscosity. Using active microrheology, the strength of elastic cages and the resulting heterogeneous distribution of mobilities have been studied in dense solutions of hard sphere colloids \cite{Habdas2004,Senbil2019}. We build on these studies and investigate recoil spectra theoretically and in simulations of (soft) repulsive particles where both approaches are known to model hard sphere colloids.

This paper is organized as follows. In Sec.~\ref{sec:mct} we present the derivation of the mode-coupling equations for general time-dependent force from which, in Sec.~\ref{sec:schematic}, we extract a schematic model whose numerical implementation is discussed in Sec.~\ref{sec:numerics}. In Sec.~\ref{sec:constant_force} we recover important results of the constant force case. Section \ref{sec:simulations} contains the description of the employed simulation method. In Sec.~\ref{sec:results} we give our analysis and comparison of both theory and simulation results for the studied recoil problem. We summarize in Sec.~\ref{sec:conclusion} and give outlook to possible future work. Certain technical aspects are included in the appendices.


\section{Mode-coupling theory}
\label{sec:mct}
\subsection{Microscopic dynamics} \vspace{-5pt}
We consider a system that consists of $N$ bath particles and one tracer, suspended in a Newtonian solvent such that an overdamped description is justified. The particle coordinates $\vr_1,...,\vr_N,\vr_s$ can be summarized as a phase space point $\Gamma$. The interaction between the particles is given by a potential $V(\Gamma)$ via $\Fi=-\del_i V(\Gamma)$, where $\Fi$ is the force acting on particle $i$. All hydrodynamic interactions are neglected. In addition the tracer particle feels a \emph{time-dependent} but homogeneous external force $\Fex(t)$. The system is in equilibrium at $t=-\infty$. The time evolution of the probability density thus follows the Smoluchowski equation \cite{Dhont}
\begin{equation}
	\partial_t\Psi(\Gamma,t) = \Omega(t) \Psi(\Gamma,t)
\end{equation}
with a time-dependent Smoluchowski operator
\begin{equation}
	\Omega(t) = \sum_{i=1}^{N,s}\left[D_i\del_i\cdot(\del_i-\beta\Fi)\right] - D_s\del_s\cdot\beta\Fex(t) ,
\end{equation}
where $D_i=k_BT/6\pi\eta a_i$ are the bare diffusion coefficients of the particles and $\beta = 1/k_BT$. The upper limit "$N,s$" of the sum means that the sum contains the $N$ terms of the bath particles and, additionally, a term for the tracer with an index $s$. For the theory we assume all particles to have the same diffusion coefficient $D_0$. The formal solution is given by \cite{Brader2012}
\begin{equation}
	\Psi(\Gamma,t) = e_+^{\int_{-\infty}^t \! ds \, \Omega(s)} \Psi_\text{eq}(\Gamma)
\end{equation}
featuring a time-ordered exponential.
Time-dependent averages of dynamical variables $A(\Gamma)$ can then be expressed via
\begin{align}
	\av{A}_t &:= \int d\Gamma \Psi(\Gamma,t) A(\Gamma) = \av{A(t)}_\eq ,
	\label{equ:av_variable}
\end{align}
having introduced a time dependence of dynamical variables
\begin{equation}
	A(\Gamma,t) := e_-^{\int_{-\infty}^t \! ds \, \Omega^\dagger(s)} A(\Gamma).
\end{equation}
The equilibrium distribution appearing in the right hand average of equation (\ref{equ:av_variable}) is assumed to be the canonical $\Psi_\eq=Z^{-1}\exp(-\beta V)$.

\subsection{Equation of motion for the transient density correlator} \vspace{-5pt}	
As will become apparent, the correlation function of interest is
\begin{equation}
	\phi_{\vk}^s(t,t') = \av{\rho_{\vk}^s,e_-^{\int_{t'}^t\! ds \,\Omega^\dagger(s)}\rho_{\vk}^s}_\eq  \equiv \av{\rho_{\vk}^s,U(t,t')\rho_{\vk}^s}_\eq,
\end{equation}
with $\rho^s_{\vk}=e^{i\vk\cdot\vr_s}$ being the Fourier transformed tracer density. To review some basic properties of this object the reader is referred to Appendix \ref{sec:prop_corr}. One has to note that it is not equal to the autocorrelation function of the density mode $\rho_{\vk}^s$ at times $t$ and $t'$ because the average is not taken over the actual probability distribution at time $t'$ but always over the equilibrium one. This means that the real space equivalent
\begin{equation}\label{eq7}
	\phi^s(\vr,t,t') = \av{\rho^s(0),e_-^{\int_{t'}^t\! ds \,\Omega^\dagger(s)}\rho^s(\vr)}
\end{equation}
is the self-part of the van Hove function only when $\Psi(\Gamma,t')=\Psi_\eq(\Gamma)$. For ease of notation, the subscript $\eq$ is dropped in Eq.~\eqref{eq7} and in the following.

We now want to derive an equation for the time evolution of $\phi_{\vk}^s(t,t')$,
which means determining the right hand side of
\begin{equation}
	\partial_t \phi_{\vk}^s(t,t') = \av{\rho_{\vk}^s,\partial_t U(t,t')\rho_{\vk}^s}.
\end{equation}	
To do this we can use the simple projection onto a mode $\rho^s_{\vk}$,
\begin{equation}
	\prop_s = \rho_{\vk}^s \av{\rho_{\vk}^s,\cdot}, ~~~~ \proq_s = 1-\prop_s.
\end{equation}
Following standard steps of Zwanzig-Mori projection formalism (see Appendix \ref{sec:derivation_of_eom}) we obtain an equation  analogous to constant force microrheology \cite{Gruber2016,GruberDiss}
\begin{align}
	\partial_t \phi_{\vk}^s(t,t') + \Gamma_{\vk}(t)\phi_{\vk}^s(t,t') + \int_{t'}^t ds' M_{\vk}(t,s')\phi_{\vk}^s(s',t')  = 0 ,
\end{align}
with the initial decay frequency
\begin{align}
	\Gamma_{\vk}(t) = D_0\left(k^2 - i \vk \cdot \beta\Fex(t)\right) ,
\end{align}
and the memory kernel
\begin{align}
	M_{\vk}(t,s') =  - \av{\rho_{\vk}^s,\Omega^\dagger(s')e_-^{\int_{s'}^tds\proq_s\Omega^\dagger(s)}\proq_s\Omega^\dagger(t)\rho_{\vk}^s} ,
\end{align}
which is also called a \emph{mobility} kernel since its Markovian approximation, viz.~assuming it fast compared to the correlator and pulling the latter out of the integral, adds to $\Gamma$, which contains the diffusion coefficient. Note the generalization of the previous equations to a dependence on two times because the present situation is not invariant under time translation.

\subsection{MCT with parallel relaxation channels} \vspace{-5pt}
To make progress one has at some point to introduce approximations of the memory kernel. Experience has shown that before doing so it is advisable to first invert the equation of motion and to introduce a friction kernel \cite{CichockiHess,Kawasaki1995,Vogel2020}. In previous work the mobility kernel has first been decomposed into the different spatial directions \cite{Gruber2016}
\begin{equation}
	M_{\vk}(t,s') = - \sum_{\alpha,\beta} L_\alpha^*(s') \mathcal{M}_{\vk}^{\alpha \beta}(t,s') R_\beta
\end{equation}
with 
\begin{gather}
	\mathcal{M}_{\vk}^{\alpha \beta}(t,s') 
	= \av{\mathcal{F}_{\vk}^\alpha,e_-^{\int_{s'}^tds\proq_s\Omega^\dagger(s)\proq_s}\mathcal{F}_{\vk}^\beta} \\
    \bm{L}_{\vk}^*(t) = D_0(\vk-i\beta \Fex(t)) \text{, } \bm{R}_{\vk} = \vk \text{, } \\
    \mathcal{F}_{\vk}^\alpha = \proq_s F_s^\alpha \rho_{\vk}.
\end{gather}
The motivation for this consideration of parallel relaxation channels was taken from MCT work on confined fluids \cite{lang_2010,lang_2013}, as velocity fluctuations tangential and perpendicular to the force contribute differently to the  mobility.
Following Gruber et al.~\cite{Gruber2016,GruberDiss} we define the operator $\Omega_\irr(s)$ via
\begin{equation}
	\proq_s\Omega^\dagger(s)\proq_s = \Omega^{\dagger}_\irr(s) - \sum_\gamma \mathcal{F}_{\vk}^\gamma \av{\mathcal{F}_{\vk}^\gamma,\cdot}.
\end{equation}
This tensorial description gives rise to the kernel relations
\begin{align}
	\mathcal{M}_{\vk}^{\alpha \beta}(t,s')
	= m_{\vk}^{\alpha \beta}(t,s') - \sum_\gamma \int_{s'}^t du \, m_{\vk}^{\alpha \gamma}(u,s') \mathcal{M}_{\vk}^{\gamma \beta}(t,u) 
\end{align}
with
\begin{equation}
	m_{\vk}^{\alpha \beta}(t,s') 
	= \av{\mathcal{F}_{\vk}^\alpha,e_-^{\int_{s'}^t\! ds \,\Omega^{\dagger}_\irr(s)}\mathcal{F}_{\vk}^\beta},
    \label{eq:def_of_m}
\end{equation}
which now represent a set of \emph{friction} kernels.
Since the equation of motion and the kernel relations cannot be transformed into Laplace space it is not possible to replace $\mathcal{M}(t,s')$ completely with $m(t,s')$ in an algebraic way as has been done in the previous work. Yet, the above coupled Volterra integral equations can in principle and, in practice in different asymptotic limits, be solved and thus are part of a closed theory.

\subsection{Mode-coupling approximation of the friction kernel}  \vspace{-10pt}
The last step is to perform the mode-coupling approximations on the friction kernels by projecting on tracer-bath density modes with
\begin{equation}
	P_2 = \sum_{\vk,\vq} \frac{1}{NS_q} \rho_{\vk}^s\rho_{\vq}\av{\rho_{\vk}^s\rho_{\vq},\cdot}.
\end{equation}
The friction kernels are thus
\begin{align}
	m_{\vk}^{\alpha\beta} (t,s')
    &\approx \av{P_2\mathcal{F}_{\vk}^\alpha,e_-^{\int_{s'}^tds\Omega_\irr^\dagger(s)}P_2\mathcal{F}_{\vk}^\beta} \\
	&= \sum_{\substack{\vp,\vq \\ \vp',\vq'}} \frac{1}{N^2 S_q S_{q'}} \av{\mathcal{F}_{\vk}^\alpha,\rho^s_{\vp'}\rho_{\vq'}} \\
	& \times \av{\rho^s_{\vp'}\rho_{\vq'},e_-^{\int_{s'}^tds\Omega_\irr^\dagger(s)}\rho^s_{\vp}\rho_{\vq}} 
	\av{\rho^s_{\vp}\rho_{\vq},\mathcal{F}_{\vk}^\beta}. \nonumber
\end{align}	
While 
\begin{equation}
	\av{\rho^s_{\vp}\rho_{\vq},\mathcal{F}_{\vk}^\beta} = i D_0 (p^\beta- k^\beta) \delta_{\vp-\vk,\vq} S_q^s
\end{equation}
can be calculated exactly, another approximation needs to be made for the four-point correlator,
\begin{align}
	&\,\av{\rho^s_{\vp'}\rho_{\vq'},e_-^{\int_{s'}^tds\Omega_\irr^\dagger(s)}\rho^s_{\vp}\rho_{\vq}} \\
	\approx&\, \av{\rho^s_{\vp'}, e_-^{\int_{s'}^t ds \Omega^\dagger(s)} \rho^s_{\vp}}
	\av{\rho_{\vq'}, e_-^{\int_{s'}^t ds \Omega^\dagger(s)} \rho_{\vq}} \\
	=&~ \delta_{\vp,\vp'} \delta_{\vq,\vq'} \phi_{\vp}^s(t,s') N S_q \phi_{\vq}(t,s'),
\end{align}
consisting of a splitting it into two-point correlators while replacing the irreducible by the full time evolution \cite{GoetzeBook}. Putting everything together and eliminating the Dirac-Deltas gives
\begin{align}
	m_{\vk}^{\alpha\beta} (t,s') \approx \sum_{\vp+\vq=\vk} 
	q^\alpha q^\beta \frac{(D_0 S_q^s)^2}{N S_q} 
    \phi_{\vp}^s(t,s')\phi_{\vq}(t,s')
\end{align}
which in the thermodynamic limit is 
\begin{align}	
	m_{\vk}^{\alpha\beta}(t,s') 
	\approx \int \frac{d\bm{p}}{(2\pi)^3} q^\alpha q^\beta \frac{(D_0 S_q^s)^2}{nS_q} \phi_{\vp}^s(t,s')\phi_{\vq}(t,s')
\end{align}
with $\bm{q}=\vk-\vp$. Inputs to this theory of $\phi_{\vk}^s(t,t')$ are the equilibrium structure factors and the bath correlator which is replaced by $\phi_{q}(t-s')$ from quiescent MCT \cite{GoetzeBook}.

\subsection{Mean (squared) displacement} \vspace{-5pt}
Although the developed theory is more general we want to restrict to the case of the force having a constant direction, which we fix to be the $z$-axis.
If one knows the tracer density correlator in Fourier space, one is also able to describe the movement of the tracer in real space since
\begin{align}
	\phi^s_{k\hat{\bm{e}}_z}(t,0) &= \av{e^{-ikz_s}e_-^{\int_0^tds\Omega^\dagger(s)}e^{ikz_s}} \\	
	&= 1 + ik\av{z_s(t)-z_s(0)} \nonumber \\
    & ~~~~~~ - \frac{k^2}{2}\av{(z_s(t)-z_s(0))^2} + ... \nonumber
\end{align}
if the system was still in equilibrium at $t=0$. This means especially that the mean and mean squared displacements (assume $z_s(0)=0$) are given by
\begin{align}
	&\av{z}_t = \bigl. -i \partial_{k_z} \phi_{\vk}^s(t,0)\bigr|_{\vk=0}, ~~~~~~ \\
	&\av{z^2}_t = \bigl. - \partial_{k_z}^2 \phi_{\vk}^s(t,0)\bigr|_{\vk=0}.
\end{align}
Through calculating the $\vk$-derivative of the equation of motion and evaluating all occuring terms for $\vk=0$ one obtains ($\Gamma_0=D_0/k_BT$)
\begin{equation} \label{eq32}
    \partial_t\av{z}_t = \Gamma_0 F_\ex(t) 
    - \int_0^t ds F_\ex(s) \mathcal{M}_0^{zz}.
\end{equation}
For $\vk=0$ we have from Eq.~(\ref{eq:def_of_m}) that $m_0^{\alpha z}=0$ for $\alpha\neq z$, such that the kernel relation is
\begin{align}
	\mathcal{M}_{0}^{zz}(t,s')
	= m_{0}^{zz}(t,s') -  \int_{s'}^t du \, \mathcal{M}_{0}^{zz}(t,u) m_{0}^{zz}(u,s') 
\end{align}
Since this equation is the statement, that $-m_0^{zz}$ is the Volterra resolvent \cite{Tricomi} of $\mathcal{M}_0^{zz}$ we can transform this into
\begin{equation}
	\partial_t\av{z}_t + \int_0^t ds' m_0^{zz}(t,s')\partial_{s'}\av{z}_{s'} = \Gamma_0 F_\ex(t)
\end{equation}
where the kernel in mode coupling approximation is
\begin{equation}
	m_0^{zz}(t,s') = \sum_{\vk} \frac{(S_k^s)^2}{N S_k} (k^z)^2 \phi_{\vk}(t,s') \phi_{\vk}^s(t,s').
\end{equation}
We also give the equation for the MSD of the quiescent system
\begin{equation}
	\partial_t \av{z^2}^\eq_t = 2D_0 - \int_0^t ds' m_z^\eq(t-s')\partial_{s'} \av{z^2}^\eq_{s'},
\end{equation}
where $m_z^\eq$ is the equilibrium (zero-force) limit of the above $m_0^{zz}$, thus containing only one time argument.
The equations for the mean displacement and the equilibrium MSD are closely linked as was elaborated in more detail in our work on the linear response case \cite{Caspers2023}.


\section{Schematic Models}
\label{sec:schematic}
It has been a common practice to employ a reduced version of the theory consisting of a system of only a few equations. This not only is of practical help in finding qualitative and analytical results and also greatly improves numerical performance. It has also been shown that  schematic models capture the bifurcation at the glass transition correctly \cite{GoetzeBook}. In the present case of the de-localization transition under force, the bifurcation has been shown to be  continuous with codimension one \cite{Gazuz2013,Gruber2020}.

\subsection{The F12 model}
The bath correlator $\phi_b$ that will go into our schematic models and describes the equilibrium dynamics of the host liquid is itself a solution of the widely employed schematic equation \cite{GoetzeBook}
\begin{equation}
	\partial_t \phi_b(t) + \phi_b(t) + \int_0^t m(t-t')\partial_{t'}\phi_b(t') dt' = 0 ,
\end{equation}
with polynomial memory kernel
\begin{equation}
	m_b = v_1 \phi_b + v_2 \phi_b^2. 
\end{equation}
In this model a line of glass transition points $(v_1^c,v_2^c)$ exists. To fix these the literature, e.g. \cite{GoetzeBook} often chooses the pair $v_2^c=2$ and $v_1^c=2(\sqrt{2}-1)$ which implies a long time correlator limit $f^c=\frac{2-\sqrt{2}}{2}$. This is taken as the reference point while the considered state is described by a distance parameter $\epsilon$ via, for example, $(v_1,v_2)=(v_1^c,v_2^c)(1+\epsilon)$. A positive $\epsilon$ produces a glass while negative $\epsilon$ a liquid. These values have been shown to produce a behaviour that also quantitatively resembles actual hard sphere systems around the glass transition.

\subsection{Derivation of a schematic model for microrheology}
To derive the schematic model for microrheology with a time-dependent force, we start from the full MCT  above and consider just two wave vectors parallel to the force with opposite sign, i.e. $\vk_\parallel$ and $-\vk_\parallel$. 
This is the minimal model that contains complex valued tracer correlators, featuring the correct bifurcation at the depinning transition and able to render real-valued observables, like the mean tracer displacement \cite{Gazuz2013}.

First we have the equation of motion for $\vk_\parallel$ (the one for $-\vk_\parallel$ is its complex conjugate)
\begin{align}
	\partial_t \phi_\parallel^s(t,t') + \Gamma_\parallel(t)\phi_\parallel^s(t,t') + \int_{t'}^t ds' M_\parallel(t,s')\phi_\parallel^s(s',t')  = 0
\end{align}
with kernel
\begin{align}
    M_\parallel(t,s') 
    = -(k_\parallel^2 - i k_\parallel F_\ex(s')) 
    \mathcal{M}_\parallel^{zz}(t,s').
\end{align} 
Because the restriction of the summation over the wave vectors (with $\vq=\vk_\parallel-\vp$) gives
\begin{equation}
    m^{xz}_\parallel (t,s') \approx \sum_{\vp=\pm \vk_\parallel} 
	q^x q^z \frac{(S_q^s)^2}{n S_q} 
    \phi_{\vp}^s(t,s')\phi_{\vq}(t,s')
    = 0.
\end{equation}
and
\begin{align}
	m^{zz}_\parallel (t,s') 
    &\approx \sum_{\vp=\pm \vk_\parallel} 
	q^z q^z \frac{(S_q^s)^2}{n S_q} 
    \phi_{\vp}^s(t,s')\phi_{\vq}(t,s')\\
    &= \frac{(S_{2k}^s)^2}{n S_{2k}} (2k)^2 \phi_{-\vk_\parallel}^s(t,s')\phi_{2\vk}(t,s'). 
\end{align}
the connection between mobility and friction kernel is
\begin{align}
    \mathcal{M}_\parallel^{zz}(t,s') 
	= m_\parallel^{zz}(t,s') - \int_{s'}^t du \, \mathcal{M}_\parallel^{zz}(t,u) m_\parallel^{zz}(u,s'). 
    \label{eq:kr_schem}
\end{align}

Since the magnitude of $\vk$ now is subsumed in a schematic vertex parameter,  $v_s$, we can write down the following dimensionless schematic equation (we also simplified the notation) 
\begin{align}
    &\partial_t \phi_\parallel(t,t') + (1-iF_\ex(t))\phi_\parallel(t,t') \nonumber \\
    & - \int_{t'}^t ds' (1-iF_\ex(s')) \mathcal{M}_\parallel(t,s') \phi_\parallel(s',t')  = 0 
\end{align}
with $m_\parallel (t,s') = v_s \phi_\parallel(t,s')^* \phi_b(t-s')$, where the bath correlator is described by the F12 model. Note that also the time at this point is dimensionless.

Now using the theory of Volterra Integral equations \cite{Tricomi} we would like to eliminate the mobility kernel from the system.
We first look at the equation of motion (EOM) rearranged to (again $\Gamma_\parallel(t)=(1-iF_\ex(t))$)
\begin{align}
	- \partial_t \phi_\parallel(t,t') 
	= \Gamma_\parallel(t)&\phi_\parallel(t,t') \\
 &- \int_{t'}^t ds' \mathcal{M}_\parallel(t,s')\Gamma_\parallel(s')\phi_\parallel(s',t'). \nonumber
\end{align}
Solving for $\Gamma_\parallel(t)\phi_\parallel(t,t')$ this results in
\begin{align}
	\Gamma_\parallel(t)\phi_\parallel(t,t') 
    = -\partial_t &\phi_\parallel(t,t')  \\
	&- \int_{t'}^t ds' H_\parallel(t,s')(-)\partial_{s'} \phi_\parallel(s',t') \nonumber
\end{align}
with a new kernel $H$ that must fulfill
\begin{align}
	H_\parallel(t,s') + \mathcal{M}_\parallel(t,s') = \int_{s'}^t du \, \mathcal{M}_\parallel(t,u)H_\parallel(u,s').
\end{align}
We can see that because of the mobility/friction kernel relation (\ref{eq:kr_schem})
\begin{align}
	H_\parallel(t,s') = - m_\parallel(t,s')
\end{align}
This results in the new EOM
\begin{align}
	&\partial_t \phi_\parallel^s(t,t')  + \Gamma_\parallel(t)\phi_\parallel^s(t,t') \nonumber \\
	&~~~~~~~+ \int_{t'}^t ds' m_\parallel(t,s')\partial_{s'} \phi_\parallel^s(s',t') = 0.
\end{align}
which is our final result for the schematic model of the tracer correlator.
Inserting a constant force will reproduce the schematic model of Gazuz et al.~\cite{Gazuz2013} for a correlator with time translational invariance. 

To extract some equivalent of average tracer motion from the schematic correlator we apply the method presented by Harrer et al.~\cite{Harrer2012} that transfers the restriction of the wave vector sum in the MCT memory kernel to the mobility kernel at $q=0$. It appears in the EOM of the mean displacement of the full theory. This means for the kernel featured in Eq.~\eqref{eq32} we get the schematic version
\begin{align}
	m_z(t,s') = &~ \sum_{\vk} \frac{(S_k^s)^2}{N S_k} (k^z)^2 \phi_{\vk}(t,s') \phi_{\vk}^s(t,s') \nonumber \\
	\approx &~ \frac{(S_k^s)^2}{N S_k} k^2 \phi_k(t,s') \cdot 2 \text{Re}\,\phi^s_k(t,s') \nonumber \\
	\equiv &~ \mu \phi_k(t,s') \text{Re}\,\phi^s_k(t,s').
\end{align}
with additional vertex parameter $\mu$. $v_s$ and $\mu$, which in the microscopic theory are given by the equilibrium structure, can in principle be used as variable fitting parameters, although in all our numerical calculations we fixed them to $v_s=4$ and $\mu=1$.

\subsection{Application to recoil}
As the first application of the above theory we want to apply the time-dependent algorithm to the step force
\begin{equation}
	F_\ex(t) = \begin{cases}
		0, t<0 \\
		F_\ex, 0<t<t_s \\
		0, t>t_s
	\end{cases} 
    \label{eq:force_protocol}
\end{equation}
while varying the magnitude $F_\ex$ and the shut-off time $t_s$ to observe the recoil motion after cessation of the driving. 

For $t>t_s>t'$ the schematic model can be written with a split integral
\begin{align}
	&\partial_t \phi(t,t') + \phi(t,t') + \int_{t'}^{t_s} \! ds \, m(t,s)\partial_s \phi_F(s-t') \nonumber \\
	&~~~~+ \int_{t_s}^t \! ds \, m_\eq (t-s)\partial_s\phi(s,t') = 0
    \label{eq:split_eom}
\end{align}
where $\phi_F$ and $m_\eq$ are the solutions for a constant or vanishing external force, that can be obtained from the time-translationally invariant schematic model solved by Gazuz \cite{Gazuz2013}.

In the numerical and simulation results that follow we will denote the total mean displacement starting from the beginning of the driving at $t=0$ as $\av{z(t)}$, while the time-dependent recoil distance is written as
\begin{equation}
    \delta z(t) = \av{z(t)} - \av{z(t_s)}
\end{equation}
and the total distance or amplitude as
\begin{equation}
    A = - \delta z(t \to \infty),
    \label{eq:amplitude_definition}
\end{equation}
which makes it a positive quantity.
The equilibrium MSD we denote in the following as $\delta z^2(t)$.


\section{Numerical Implementation}
\label{sec:numerics}

The numerical method we use for the recoil problem relies on both the solution of the constant force case and thus a one-time numerical scheme and an extended scheme featuring two time-arguments. The derivation and discussion of the one-time scheme can be found in previous publications on this topic, e.g. \cite{Gruber2016}. In the beginning of the study we used a simpler two-time solution scheme, more akin to the one used by Frahsa \cite{FrahsaDiss}. This had the advantage of being quite versatile and fast but turned out to be numerically less precise than the current realisation. Nevertheless all results we could also achieve qualitatively with the former algorithm, so it was a helpful tool \cite{Ditz2024}.

The details of the numerical scheme are explained in appendix \ref{sec:numerics_details}. The idea for it was developed originally in \cite{Voigtmann2012} by Voigtmann and colleagues. It is tailored to step force protocols and makes explicit use of the fact that the equation of motion for the correlator can be written as Eq.~(\ref{eq:split_eom}). While in \cite{Voigtmann2012} the method was applied to standard rheological shear our schematic model for microrheology has a different time structure in the vertex parameters.
Overall it is a simpler application of the numerical ideas that is still challenging to implement due to the doubling of the time arguments compared to the constant force case.

To understand some notation in the main text we give a quick introduction here. Because of the two time arguments $t\geq t'$ in principle one needs to solve the equations on a triangular time grid $(i\Delt,j\Delt)$ with some small time-step $\Delt$ and $j\leq i$. All functions $f$ exist on this grid, $f_{ij}=f(i\Delt,j\Delt)$. The grid contains a number of $N_t \times N_t$ points. To greatly increase the time window that can be calculated by the method we employ a \emph{decimation} algorithm, cp. \cite{Gruber2016}. It occurred useful to make every numerical parameter a power of 2. This is because of the \emph{doubling} of the time step in every decimation window. The numerical quality, i.e. the degree of convergence of the result to the actual solution of the integro-differential equation (IDE) is best expressed by the quotient $\frac{N_t}{t_s}$ which is the density of points in the last calculation. This means we get the best results if we cessate the force as early as possible e.g. after the system has reached its steady state and increase the overall number of grid points as far as computationally feasible. 

Due to the increased complexity we can assume that the numerics contains more uncertainty than in the constant force case. Still we could see that in the two test cases of a constant force and the linear response, Sec. \ref{sec:lin_res}, the algorithm performs satisfyingly. We have also checked proper convergence in the higher force cases and can assume the results to be valid solutions of the EOM.


\section{Constant force results}
\label{sec:constant_force}
In this short section we would like to summarize some previous results on the constant force case. They are not new but we find it useful to highlight the ingredients from previous work that are the basis for our studies of the switch-off force. We also indicate the numerical challenges in the two-time algorithm that follow from these results.


\subsection{Tracer correlators and stationary state}
The F12-model enters the model for our tracer particle as an expression of the state of the bath. 
For states $\epsilon<0$, which represent a fluid, the bath correlation function decays to zero, while for $\epsilon>0$, it decays to a nonzero plateau which is called the nonergodicity parameter signalling a glassy system. 
Also there exists a region where $0<-\epsilon \ll 1$, the supercooled regime where we encounter a two-step relaxation process. For a fluid, the two relaxation processes have merged and correlators decay in a single process. 
Sufficiently away from the glass transition the correlators decay within a time-window of about $10^2-10^4$, which is where most of the following studies takes place. The long decay time has great influence on the precision of the numerical solution which is commented on more in the appendix. 

Considering now an applied constant force in the glass case of positive $\epsilon$, the most important finding has been that there exists a threshold for the force magnitude $F_\ex=F_c$ \cite{Gazuz2009,Gruber2016}. Above this critical force, the tracer particle is pulled free from its cage of neighbours, which signals a depinning transition. Analytical expressions for the critical force have been derived for the schematic model \cite{Gazuz2013,Gruber2020}. The tracer density correlator for $F_\ex>F_c$ decays completely to zero, following a power law with exponent $-1/2$.


\begin{figure}
	\includegraphics{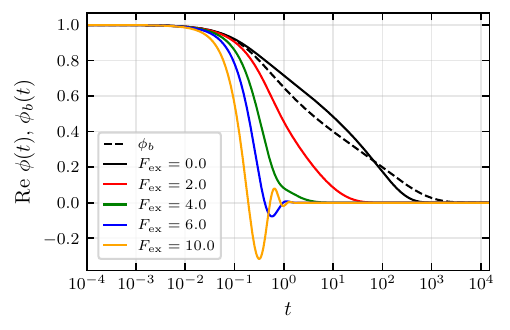}
	\caption{Real part of the tracer correlator for a liquid, $\epsilon=-0.1$. Upon increasing the external force the correlator decays faster. For higher forces there arise oscillations. The dashed line shows the bath correlator taken from the F12-model. It is very similar to the tracer correlator with zero force.}
    \label{fig:fluid_const_force}
\end{figure}

For negative $\epsilon<0$ the correlators always decay to zero for all forces, see Fig.~\ref{fig:fluid_const_force}. At some point, increasing the force leads to oscillations, which is also true for the glass case. The bath correlator is included in the figure and is similar to the tracer correlator in the limit of small force. This explains why e.g. tracer experiments can be used to gain information about the whole suspension.

The stationary velocity that is reached in the steady state is given by
\begin{equation}
	v_{st} = \frac{F_\ex}{1+\int_0^\infty dt \, m_z(t)},
\end{equation}
for any value of $\epsilon$. For subcritical forces, the integral clearly is infinite since the kernel does not decay to zero and thus $v_{st}=0$.
We see that for increasing force the oscillations in the correlator lead eventually to the vanishing of the integral meaning $v_{st}=F_\ex$ (in the schematic model there is $D_0=1$) which already states that for high 
enough forces the system is not deformed elastically (leading to a decrease of final velocity) anymore, the particle just breaks through the system. On the other hand the magnitude of this effect seems to be overestimated by the theory, since instead there should be a nontrivial stationary friction coefficient \cite{Squires2005}, as has been shown in simulations \cite{Gazuz2009}.

To prepare for our calculations of the recoil, we have examined at which time the stationary velocity is reached. This would then be a guideline to when the force should be shut off. The time agrees with the decay time of the tracer correlator. It is force-dependent and highest for low forces. 

\subsection{Analytical predictions for recoil from the constant force calculations}
\label{sec:const_pred}
Without actually calculating the recoil directly, we are able to derive some expectations about the recoil amplitude from the stationary constant force results. For this we consider the equation of motion for the mean displacement for a very small time after the force is shut off. This provides the initial backward velocity the colloid tracer experiences. If we assume continuity of the two-time kernel function $m_z$ in both arguments, also across the shut-off time, which is large, we get 
\begin{align}
    \partial_t z(t_s+\delta t)
    &= - \int_0^{t_s+\delta t} \! ds \, m_z(t_s+\delta t,s)
    \partial_s z(s) \nonumber \\
    &\approx - \int_0^{t_s} \! ds \, \, m_z(t_s,s)
    \partial_s z(s) + F_\ex - F_\ex \nonumber \\
    &= v_{st} - F_\ex \nonumber \\
    &= - F_\ex \frac{\int_0^\infty dt \, m_z(t)}{1 + \int_0^\infty dt \, m_z(t)}
\end{align}
For ease of notation we have omitted averaging brackets in these equations.
To understand this result we re-express it with actual units as
\begin{equation}
    - v_\text{init} = \frac{F_\ex}{\zeta_0} - \frac{F_\ex}{\zeta} = \frac{F_\ex}{\zeta_0} \frac{\int_0^\infty dt \, m_z(t)}{1 + \int_0^\infty dt \, m_z(t)}
\end{equation}
with $\zeta = \zeta_0(1+\int_0^\infty dt \, m_z(t))$.

\begin{figure}
	\centering
	\includegraphics{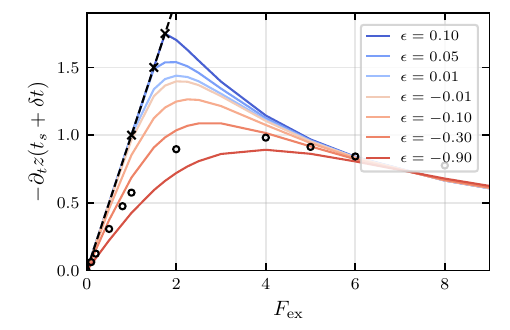}
	\caption{Backward velocity directly after switch-off of the external driving force (from approximation described in the text) for varying $\epsilon$. In the glassy state the system follows the linear curve $v=F$ until the critical force is reached. Then the curves reach a maximum which coincides with the critical force (crosses) if the system is sufficiently away from the glass transition. For decreasing $\epsilon$ the curve becomes more and more smeared-out with the maximum shifting to higher forces. This happens smoothly across the glass transition. We added (as circles) already results from the nonlinear two-time numerics for $\epsilon=-0.3$.}
	\label{fig:v_init_prediction}
\end{figure}

The numerical results, seen in Fig.~\ref{fig:v_init_prediction}, show that this expression peaks for all $\epsilon$ due to the force-dependent behaviour of $\int_0^\infty dt \, m_z(t) \sim \frac{1}{F_\ex-F_c}$. In the glass it has been found before, that close to the depinning transition $v_{st}=c(F_\ex-F_c)$ \cite{Gruber2020}.
In this case we also have a more clear separation
\begin{equation}
    - v_\text{init}
    = \begin{cases}
        F_\ex/\zeta_0, &\text{if } F_\ex < F_c \\
        F_\ex/\zeta_0 - v_{st}, &\text{if } F_\ex > F_c
    \end{cases}
\end{equation}
so above the critical force elastic and plastic deformations begin to compete leading to a peak slightly above the critical force when the system is very close to the glass transition. 
For liquid $\epsilon$ the curve has a softer peak that shifts to higher forces. We would reason that the total recoil distance $A$ (defined in Eq.~\ref{eq:amplitude_definition}) is directly related to this initial backward velocity thus we would predict a maximum of this amplitude around the same position on the force-axis.
The decrease of $v_\text{init}$ happens because in this schematic model the high-force friction coefficient is converging to the trivial $\zeta_0$, which is in contrast to simulation results \cite{Gazuz2009}. This could be a strong reason for the later occurrence of the strongly non-monotonous recoil amplitude.


\section{Simulations}
\label{sec:simulations}
Simulations of spherical particles have been run to test the theoretical predictions. $N$ particles in a cubic box with periodic boundary conditions are considered. The particles radii are drawn from a flat distribution of width $2\delta = 0.2a$, with $a$ the average radius. The particle-particle interaction potential is continuous but steep enough to give the same results as a hard core potential \cite{Lange2009}; its explicit form is:

\begin{equation}
V(r) = \kbT \left( \frac{r}{a_{ij}} \right)^{-36}
\end{equation}

\noindent with $a_{ij}=a_i + a_j$, the center-to-center distance between particles $i$ and $j$ at contact. All particles follow Langevin dynamics, i.e.~for particle $j$, the equation of motion reads:

\begin{equation}
m \frac{d^2\, {\bf r}_j}{dt^2}\:=\: \sum_{i\neq j} {\bf F}_{ij} - \gamma_0 \frac{d\, {\bf r}_j}{dt} + {\bf f}_j(t) + \Fex (t) \delta_{js}
\label{Langevin}
\end{equation}

\noindent where $m$ stands for the particle mass (identical for all particles), ${\bf F}_{ij}$ is the central force between particles $i$ and $j$ derived from the previous interaction potential, $\gamma_0$ is the friction coefficient with a solvent, ${\bf f}_j$ is a random force, that fulfills the fluctuation-dissipation theorem, and $\Fex$ is the external force acting onto the tracer. The latter has radius $a_s=a$, and is pulled with the time-dependent force given by (\ref{eq:force_protocol}). The bath density, measured as volume fraction of particles, is varied approaching the glass transition; $\varphi=0.50$, $0.55$ and $0.57$ are considered.

Different configurations were equilibrated with every density for a long time before the force is applied, and around 10000 different trajectories were studied for every case. In the simulations, standard units are used; average particle radius, $a$, particle mass, $m$, and thermal energy $\kbT$ are the units of length, mass and energy, respectively.

While Langevin dynamics is used to simulate Brownian systems, it keeps the inertial term. This inertia can be relevant when dealing with time-dependent forces, as it introduces an internal time scale due to the competition between inertia and dissipation, $m/\gamma_0$. The effect of this inertia on the tracer trajectory is studied in Appendix \ref{app_sims}. An optimal value of $\gamma_0=100\,\sqrt{m\kbT}/a$ was found, where inertia has negligible effects, while keeping the simulation time at reasonable levels.


\section{Results for recoil}
\label{sec:results}
Having now described all necessary theoretical and numerical tools and having covered some results for constant applied forces we now want to present our results for time-dependent forcing on a first simple application, namely the switch-off force protocol that leads to the recoil phenomenon. First we show that we can recover the linear response formula and assess the validity of this approximation. We follow this up by showing that also in the simulations the same linear response formula holds, even for driving in  non-stationary states. Our analysis continues with the results of our two-time-scheme calculations for higher forces where we see that the schematic model predicts a non-monotonous recoil amplitude. We finally perform a mapping of our schematic model to the simulation data which show similar behaviour concerning the crossover into nonlinearity but no clear corroboration of the predicted phenomenon at increasing forces.

\begin{figure}
	\centering
	\includegraphics{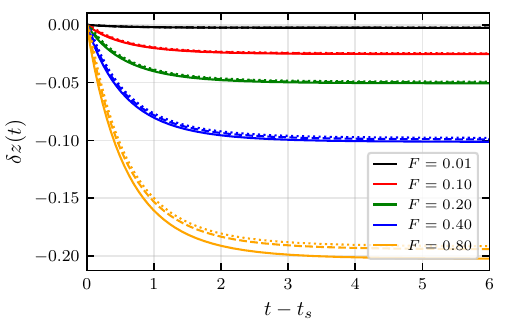}
	\caption{For same colors these curves show time-dependent recoils for $N_t=4096$ and $N_t=2048$ at $\epsilon=-0.9$ as dashed lines compared to the result from applying the recoil formula to the equilibrium MSD. For each color the order of the curves is from top to bottom: 2048, 4096, LR formula. $t_s=64$ in this case.}
	\label{fig:linear_response_bereich}
\end{figure}

\subsection{Linear response}
\label{sec:lin_res}
It is not hard to show analytically that the schematic model fulfills the linear response formula for the time-dependent recoil, cp. \cite{Caspers2023}, in the general form
\begin{equation}
    -\langle \delta z(t) \rangle = \frac{F_\ex}{2} \lbrace
    \langle \delta z^2 (t-t_s) \rangle + \langle \delta z^2 (t_s) \rangle -
    \langle \delta z^2 (t) \rangle \rbrace,
    \label{caspers2023b}
\end{equation}
while the formula for the force-on part is given by
\begin{equation}
    \langle z(t) \rangle = \frac{F_\ex}{2} 
    \langle \delta z^2 (t) \rangle. 
    \label{lin_resp_const}
\end{equation}
In order to test this relation numerically, it must be noted that only the r.h.s. of (\ref{caspers2023b}) can be determined quite precisely in a one-time calculation while the left hand side is calculated from the two-time scheme that inevitably contains errors that get lower by increasing $N_t$, while keeping $t_s$ constant. Nevertheless this is computationally only feasible up to a certain limit. In the cases we considered we chose the restriction of $N_t\leq8192$.

In Fig.~\ref{fig:linear_response_bereich}, recoils are shown for small forces and compared to the linear response prediction. Since our results are never fully converged numerically we plot for each force two (dashed) curves produced by calculations with $N_t=4096$ and $N_t=2048$. The linear response prediction is shown as a solid line in the corresponding color. We can see (for this example of $\epsilon=-0.9$) that around $F=0.4$ the actual recoil result starts to deviate from the linear response formula prediction, meaning the distance between the lowest dashed line and the solid line gets larger than the distance between the two dashed lines. The forcing time $t_s=64$ was chosen high enough to get into the steady state. It can be concluded from this figure that linear response is valid for small forces.

\begin{figure}
	\includegraphics{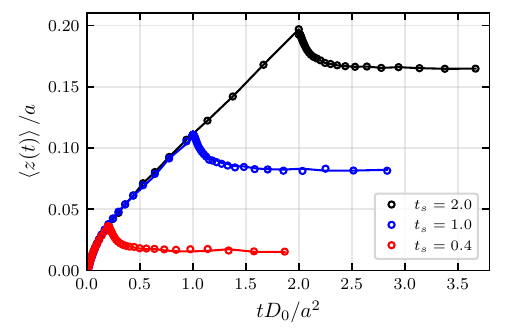}
	\includegraphics{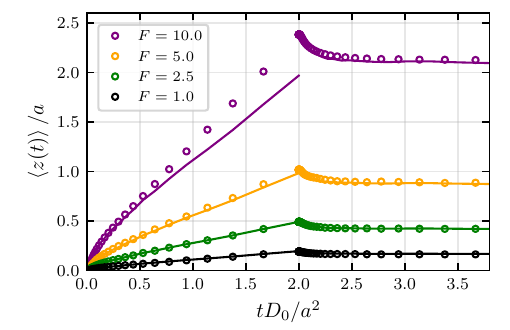}
	\caption{Comparison of the tracer displacement from simulations (circles) with linear response theory, for different driving times  (upper panel at $F_\ex=1$) and  forces (lower panel, $t_s=2$). In all cases, the volume fraction is $\varphi=0.50$, and the friction coefficient is $\gamma_0 = 5 \sqrt{m\kbT}a$. Note, the black curves agree in both panels.}
	\label{fig:recoil_sim_LR}
\end{figure}

Typical curves of the tracer displacement in the force direction from simulations are shown in Fig. \ref{fig:recoil_sim_LR} (in all cases the bath volume fraction is $\varphi=0.50$). The figure includes the forcing regime ($t<t_s$) and recoil ($t>t_s$). 
The result of the linear response formulas is included in the figure as lines.
The upper panel shows the effect of increasing the driving time, $t_s$, whereas the lower one studies different forces; in both cases, the pulling and non-pulling regimes are studied independently, introducing an offset at $t_s$ to match the tracer displacement between simulations and theory. During the driving, a transient regime is observed for short times, which crosses over to a long-time steady state, characterized by a constant velocity. The LR calculation agrees for small forces but deviates systematically for large ones, in the regime $t<t_s$ but also in the recovery back to equilibrium $t>t_s$. This indicates the limit of the linear regime. Notably, the agreement between simulations and LR theory is observed irrespective of the driving time being shorter or longer than this transient regime (upper panel of the figure).

\subsection{Non-linear theory results}
\begin{figure}
	\centering
	\includegraphics{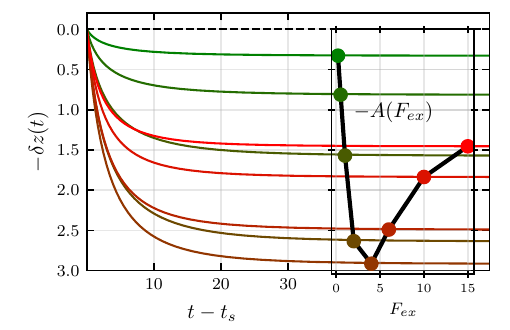}
	\caption{Main plot: Recoil curves at $\epsilon=-0.9$ with $t_s=64$ and $N_t=8192$. The inset/subplot at the right side illustrates the nonmonotonous behaviour of the curves with increasing force. Color code shows transition from low forces (green) to high forces (red). But the mapping of curves to its corresponding force magnitude is also given by the x-axis of the force plot.}
	\label{fig:recoils_bis_15}
\end{figure}

If we now move on to consider the region of higher forces we notice two things (Fig. \ref{fig:recoils_bis_15}). First the shape of the recoil curves remains almost unchanged being almost completely determined by the total distance that is travelled back by the particle and, second, this amplitude decreases again beyond a certain force value. This is in agreement with the prediction coming from the initial backward velocity calculated for constant force in section \ref{sec:const_pred}. We see that the force of maximum recoil could have been predicted correctly from the force of maximum initial backward velocity in Fig.~\ref{fig:v_init_prediction}.
Apparently, the amplitude of the recoil is a measure of the elastic stresses stored in the surrounding medium. They push the tracer particle back, and the asymptotic recoil amplitude varies with the initial velocity. The stored elastic stresses get   weakened by plastic rearrangements, which limits the linear growth of the recoil with the force. Interestingly, the schematic model predicts a non-monotonous behavior. For large forces, the plastic processes outgrow the elastic storage so that the recoil amplitude shrinks. 
For the chosen $\epsilon=-0.9$ we have checked that the steady state has been reached after $t_s=64$. The recoil in general saturates quite fast but the saturation time is slightly longer for curves with higher amplitude but shorter for higher force. 

\begin{figure}
	\centering
	\includegraphics{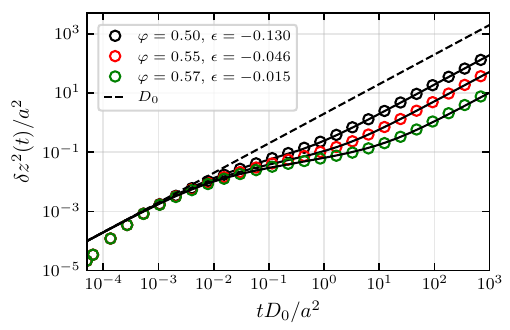}
	\caption{Fits of the schematic model equilibrium MSDs (solid black lines) to corresponding MSDs calculated from Langevin simulations (circles). For the fits only the non-ballistic part of the curves was considered. The fitted scaling factor $b$ (see text) is from top to bottom: $0.0071$, $0.0044$, $0.0032$.}
    \label{fig:fit_msd}
\end{figure}

\subsection{Comparison with simulations}
To map our schematic model to the results from the simulations we proceeded in two steps. First we want to fit the MSDs by choosing for any packing fraction $\varphi$ an $\epsilon$-value for the schematic model and then scaling $y$- and $x$-axis of the schematic MSDs such that they fit to the simulation MSDs. We have to deal with the situation that the simulations are performed in Langevin dynamics, so have a different short-time behaviour than the schematic model and only agree for longer times. 
The scaling factors used for time- and space- axes should be the same factor $b$ since in the chosen diffusion units of the plot the  short-time slope equals 1. 
Figure \ref{fig:fit_msd} shows the result of these fits for the three dense fluid packing fractions that were used in the simulations. Exemplarily, for the simulations at packing fraction $\varphi=0.5$, we found good agreement with the schematic result for $\epsilon=-0.13$ and $b=0.0071$. These are the only density dependent fit parameters we use, while all other parameters of the model are constant.

\begin{figure}
	\centering
	\includegraphics{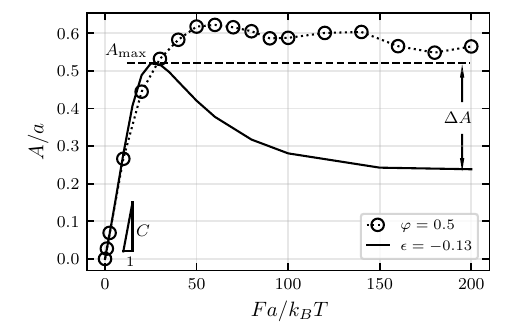}
    \caption{Recoil amplitudes $A$ as a function of external force magnitude $F$ from simulation ($\varphi=0.5$, symbols) and schematic model ($\epsilon=-0.13$, solid line). The schematic model has been mapped to the simulated system via the equilibrium MSD. The added labels in the plot illustrate the properties defined in Eq.~(\ref{eq:prop_1}-\ref{eq:prop_3}). }
    \label{fig:fit_to_amplitudes}
\end{figure}

As a second step we directly fit the recoil curves of the simulations with those from the corresponding (by means of the MSD fit) schematic model. Because of the difference in short time dynamics we do not expect agreement in the shape of the curves but more in the amplitudes.
The linear response formula holds for the simulated system up to a value of about $Fa/k_BT=5$ while for the schematic model it only holds until $F=0.5$. To accommodate for this difference we introduce a force-rescaling by a factor of 10. By this we mean that to fit the simulated recoil at $F=1$ we calculate the schematic result for $F=0.1$ but multiply it again by 10 to achieve the correct amplitude. This is also kept fixed for all densities. Thus we have fitted explicitly the linear response regime and can afterwards observe how both systems behave then for higher forces. Fig. \ref{fig:fit_to_amplitudes} shows such a fit for $\varphi=0.5$ or $\epsilon=-0.13$, respectively. One can see how the fit was made to agree especially in the initial linear response slope. It is interesting to note that the maximum that is reached shortly after the linear regime still agrees quite well, even though the position from theory is somewhat lower on the force-axis and in height. The decay after the maximum differs. Theory predicts a rather strong reduction of the recoil, while simulation finds a much smaller effect. This behavior might in part be traced back to the observation that the schematic model  underestimates the tracer friction coefficient at higher forces \cite{Gazuz2013}.

\begin{figure}
	\centering
	\includegraphics{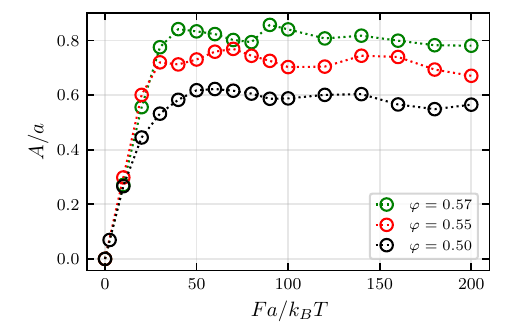}
	\caption{Recoil amplitude from simulations for different volume fractions, as labeled. \label{recoil_sim_phi}}
\end{figure}

Fig. \ref{recoil_sim_phi} shows the simulation results of the recoil amplitude for increasing densities, deeper in the viscoelatic regime.
The figure shows slight changes in the linear response region, but more important changes for larger forces. Here, the recoil amplitudes grow with packing fraction. The shape of the curves does not strongly change, and for all $\varphi$ a maximum, as clear as found in the theory, see solid line in Fig. \ref{fig:fit_to_amplitudes}, is not observed.

To summarize the comparison between schematic model theory and simulations, we define three properties to describe the force-dependent recoil $A(F)$-curves. These are the initial slope, the maximum value and the difference to the high-force limit:  
\begin{gather}
    C \equiv \lim_{F\to 0} \frac{\partial A(F)}{\partial F} \label{eq:prop_1}  \\
    A_\text{max} \equiv \text{max}_F A(F) \\
    \Delta A \equiv A_\text{max} - A(F\to \infty)  \label{eq:prop_3}
\end{gather}

\begin{figure}
	\centering
	\includegraphics{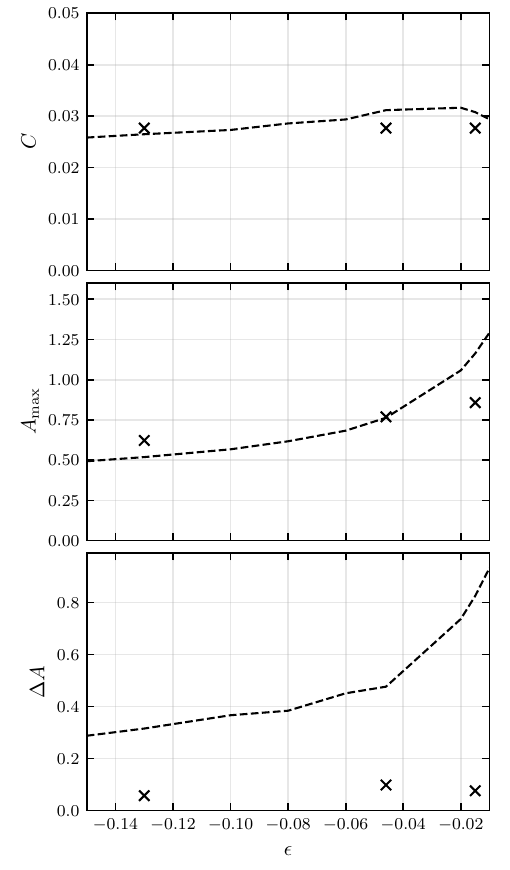}
 \caption{Properties of the force-amplitude curves. The three panels show from top to bottom the initial slope $C$ of the function $A(F)$ for low forces, the maximum height of the curve $A_\text{max}$ and the difference between the maximum and the value for very high forces $\Delta A$. The points are the results from the simulations while the dashed lines respresent the theory results using the mapping that was established via the comparison of MSDs, while using interpolation of the fit parameters for densities not covered in the simulations.}
    \label{fig:properties_of_amplitudes}
\end{figure}

The initial slope $C$ can be interpreted as a compliance, because it links the distance the tracer is pushed back by the stored elasticity to the strength of the forcing before letting the tracer go, viz.~switching off the force. The maximal recoil amplitude clearly is a measure of the maximal elastic deformation the tracer can impress on the viscoelastic fluid. The decrease of the recoil amplitude from its maximum to the large force limit ($\Delta A$) measures the magnitude of plastic effects.
These quantities are  illustrated in Fig. \ref{fig:fit_to_amplitudes}.
To compare the results we interpolate the mapping we obtained for the three values of the packing fraction $\varphi$ above to cover a more continuous $\epsilon$-range. We used
$b(\epsilon) = 0.0071 - 0.0339(\epsilon+0.13)$. This is shown in Fig.~\ref{fig:properties_of_amplitudes}. 

We observe that this mapping results in an overall mostly constant initial slope over the range of considered values. This can be seen as a typical result of MCT, where glassy properties are asymptotically constant in the fluid state.

Also the increase in the maximum amplitude is captured, albeit somewhat exaggerated by the theory. As stated before the main difference is the behaviour for high forces. There, the recoil amplitude decreases with increasing force, viz.~ $\Delta A$ is positive, but the effect is far smaller in the simulation than predicted. Still the simulations all show a (broad) maximum  in the intermediate force range and a slight downward trend for increasing force.


\section{Summary and Conclusions}
\label{sec:conclusion}
In this work we have extended the framework of MCT microrheology to time-dependent forces and evaluated it for the case of a step force within the schematic model approach. Additionally we have performed Langevin simulations of polydisperse hard spheres to assess the validity of the theoretical approximations.

Once more, in addition to the previous publication \cite{Caspers2023} the linear response recoil formula has been verified for theory and simulations. On the other hand the comparison of the nonlinear displacement-force regime has provided additional insights. Firstly, the linear regime is quickly followed by a  maximum in the recoil amplitude $A$. Nonlinear elastic effects are rather small. Rather plastic phenomena quickly set in and delimit $A$. For large forces, the recoil approaches a finite amplitude. The schematic model predicts a clear maximum for intermediate forces, which is not as clearly visible in the simulation. 
The schematic model seems to overestimate the destruction of stresses in front of the tracer particle for the higher forces which at this point can be seen as the main reason for the strong non-monotonicity of the recoil amplitude. This deficiency of MCT to capture the (rather) high friction in the strong-force case \cite{Squires2005} was already previously noted \cite{Gnann2011}. This overestimate  is also clearly observed for the initial velocity after force shut-off. There might be potential for adjustments to the model such that e.g. a non-trivial friction coefficient for (arbitrarily) large force is reproduced, which could potentially significantly weaken the observed $\Delta A$. 

One motivation for this work was to find out if there is a remnant of the critical force even in the fluid state. This seems to be the case for the schematic model although at this point results of either simulation or numerics very close or beyond the glass transition have not yet been achieved. For the presented simulation results we have found that the maximum or plateau in the recoil is reached for force magnitudes that are similar to the critical force in previous simulations.

As the theory can be generally applied to any time-dependent force it should be tested for more cases in future work. One possibility would be to use it to numerically find $F_\ex(t)$ such that  a constant mean velocity can be regulated. This would allow to study e.g. the differences of the  distribution functions in the two cases, constant velocity and  constant force.

Also developing a schematic model that incorporates a perpendicular wave vector as in Ref.~\cite{Gruber2020} could be used to study inter alia the perpendicular motion during the recoil.


\section*{Acknowledgements}
 N.D. and M.F. acknowledge support from project C6 of SFB1432 and thank M. Krüger, C. Bechinger, A. Zumbusch, and all colleagues from projects C5 and C7 for discussions. A.M.P. acknowledges financial support
through project No. PID2021-127836NBI00 (funded by MCIN/AEI/10.13039/501100011033/FEDER “A way to make Europe”).\\

\newpage
\appendix

\section{Properties of the tracer correlator} \label{sec:prop_corr}
In case of translation invariance
\begin{align}
	\phi^s_{\vk,\vk'}(t,t') =& \av{e^{-i\vk\cdot\vr_s},e_-^{\int_{t'}^tds\Omega^\dagger(s)}e^{i\vk'\cdot\vr_s}} \nonumber \\
	\stackrel{TI}{=}& \av{e^{-i\vk\cdot(\vr_s+\bm{a})},e_-^{\int_{t'}^tds\Omega^\dagger(s)}e^{i\vk'\cdot(\vr_s+\bm{a})}} \\
	=& \av{e^{-i\vk\cdot\vr_s},e_-^{\int_{t'}^tds\Omega^\dagger(s)}e^{i\vk'\cdot\vr_s}}e^{-i\bm{a}\cdot(\vk-\vk')}, \nonumber
\end{align}
which implies that $\phi^s_{\vk,\vk'}(t,t')$ can only be nonzero if $\vk=\vk'$.

The rotational symmetry around the axis of the external force can be shown similarly to the constant force case \cite{GruberDiss}. To do it we look at the equation of motion
\begin{align}
	&\partial_t \phi_{\vk}^s(t,t')  + \Gamma_{\vk}(t)\phi_{\vk}^s(t,t') \\
	&~~~~+ \int_{t'}^t ds' M_{\vk}(t,s') \phi_{\vk}^s(s',t') = 0.
\end{align}
We want to show that $\phi_{\mathcal{R}\vk}^s = \phi_{\vk}^s$ if $\mathcal{R}$ is some rotation around the external force axis. This needs to hold if $\Gamma_{\vk}$ and $M_{\vk}$ are invariant under this rotation since then both correlators fulfill the same differential equation. Those two quantities can be expressed as (using vector and matrix multiplication)
\begin{align}
	&\Gamma_{\vk}(t) = \bm{L}_{\vk}^*(t)^T \bm{R}, \\
	&M_{\vk}(t,s') = \bm{L}_{\vk}^*(s')^T \bm{\mathcal{M}}_{\vk}(t,s') \bm{R}_{\vk} .
\end{align}
with
\begin{equation}
    \bm{L}_{\vk}^*(t) = \vk-i\beta \Fex ~~~ \text{and} ~~~
    \bm{R}_{\vk} = \vk
\end{equation}
From Gruber \cite{GruberDiss} (Chapter 2.4.4) we take that
\begin{align}
	&\bm{L}_{\mathcal{R}\vk}^*(s')^T = \bm{L}_{\vk}^*(s')^T \mathcal{R}^T, \\
	&\bm{R}_{\mathcal{R}\vk} = \mathcal{R}\bm{R}_{\vk}, \\
	&\bm{\mathcal{M}}_{\mathcal{R}\vk}(t,s') = \mathcal{R} \bm{\mathcal{M}}_{\vk}(t,s') \mathcal{R}^T,
\end{align}
which shows (using $\mathcal{R} \mathcal{R}^T = 1 $)
\begin{equation}
	\Gamma_{\mathcal{R}\vk} = \Gamma_{\vk}, ~~~ M_{\mathcal{R}\vk} = M_{\vk}.
\end{equation}

\section{Derivation of the equation of motion} \label{sec:derivation_of_eom}
By splitting the adjoint Smoluchowski operator like $\Omega^\dagger(s)=\prop_s\Omega^\dagger(s)+\proq_s\Omega^\dagger(s)$ and using an operator identity we obtain
\begin{widetext}
\begin{align}
	U(t,t') &= e_-^{\int_{t'}^tds\Omega^\dagger(s)} = e_-^{\int_{t'}^tds\prop_s\Omega^\dagger(s) + \proq_s\Omega^\dagger(s)}\\
	&= e_-^{\int_{t'}^tds\proq_s\Omega^\dagger(s)} + \int_{t'}^tds' e_-^{\int_{t'}^{s'}ds\Omega^\dagger(s)} \prop_s \Omega^\dagger(s')
	e_-^{\int_{s'}^tds\proq_s\Omega^\dagger(s)}.
\end{align}
The derivative of this operator with respect to $t$ is
\begin{align}
	\partial_t U(t,t') = e_-^{\int_{t'}^tds\proq_s\Omega^\dagger(s)}&\proq_s\Omega^\dagger(t) + e_-^{\int_{t'}^tds\Omega^\dagger(s)}\prop_s\Omega^\dagger(t)\\
	& + \int_{t'}^tds' e_-^{\int_{t'}^{s'}ds\Omega^\dagger(s)} \prop_s \Omega^\dagger(s')
	e_-^{\int_{s'}^tds\proq_s\Omega^\dagger(s)}\proq_s\Omega^\dagger(t).
\end{align}
The inner products of these three terms with the tracer density mode are
\begin{align}
	\av{\rho_{\vk}^s,e_-^{\int_{t'}^tds\proq_s\Omega^\dagger(s)}\proq_s\Omega^\dagger(t)\rho_{\vk}^s} 
	&= \av{\rho_{\vk}^s,\proq_s e_-^{\int_{t'}^tds\proq_s\Omega^\dagger(s)}\proq_s\Omega^\dagger(t)\rho_{\vk}^s}\\
	&= \av{\proq_s\rho_{\vk}^s, e_-^{\int_{t'}^tds\proq_s\Omega^\dagger(s)}\proq_s\Omega^\dagger(t)\rho_{\vk}^s}\\
	&= \av{0, e_-^{\int_{t'}^tds\proq_s\Omega^\dagger(s)}\proq_s\Omega^\dagger(t)\rho_{\vk}^s}\\
	&= 0,
\end{align}
\begin{align}
	\av{\rho_{\vk}^s,e_-^{\int_{t'}^tds\Omega^\dagger(s)}\prop_s\Omega^\dagger(t)\rho_{\vk}^s}
	= \av{\rho_{\vk}^s,U(t,t')\rho_{\vk}^s}\av{\rho_{\vk}^s,\Omega^\dagger(t)\rho_{\vk}^s} 
	\equiv - \phi_{\vk}^s(t,t') \Gamma_{\vk}(t)
\end{align}
and
\begin{align}
	&\av{\rho_{\vk}^s, \int_{t'}^tds' e_-^{\int_{t'}^{s'}ds\Omega^\dagger(s)} \prop_s \Omega^\dagger(s')
	e_-^{\int_{s'}^tds\proq_s\Omega^\dagger(s)}\proq_s\Omega^\dagger(t)\rho_{\vk}^s}\\
	=& \int_{t'}^tds' \av{\rho_{\vk}^s,U(s',t')\rho_{\vk}^s} \av{\rho_{\vk}^s,\Omega^\dagger(s')
	e_-^{\int_{s'}^tds\proq_s\Omega^\dagger(s)}\proq_s\Omega^\dagger(t)\rho_{\vk}^s}\\
	\equiv& -\int_{t'}^tds' \phi_{\vk}^s(s',t') M_{\vk}(t,s').
\end{align}
\end{widetext}

\section{Details On Numerical Implementation}
\label{sec:numerics_details}
\begin{figure}
	\centering
	\includegraphics[width=0.40\textwidth]{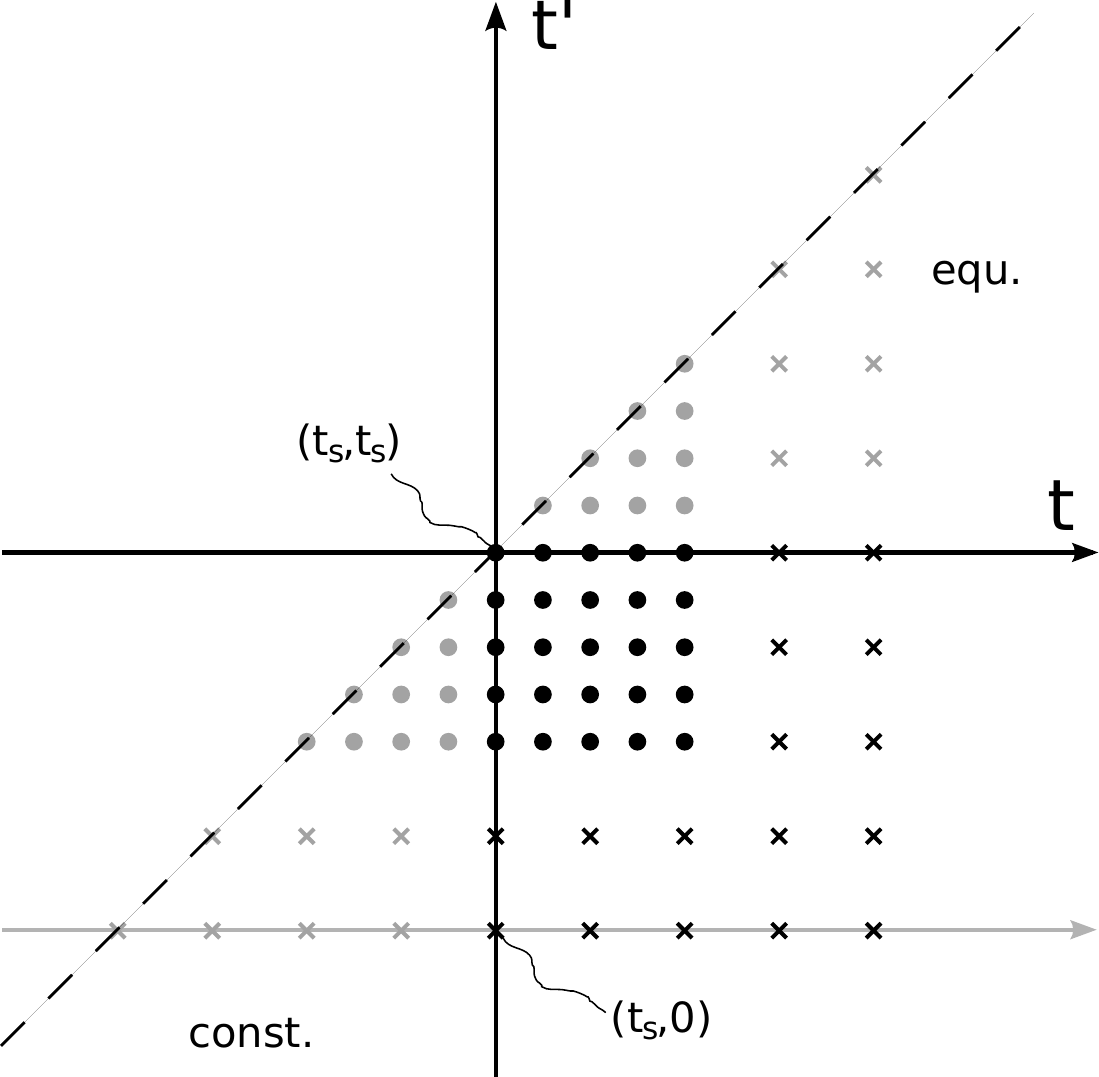}
	\caption{Illustration of the two-time numerical scheme. The black circles represent the grid points to be calculated in a first step, while the grey circles are calculated from the one-time schemes. After decimation additional grid points (crosses) are calculated until $t_s$ is reached. Usually this is repeated far more often than once, in contrast to what is depicted due to simplicity.
    In the notation described in the text, here $N_t=4$ and $n=1$.}
	\label{fig:voigtmann_scheme}
\end{figure}

Here we present the numerical scheme that was used for the solution of the equation of motion of the schematic model and corresponding mean displacement. It is also illustrated in Figure \ref{fig:voigtmann_scheme}.

Because of the two time arguments $t\geq t'$ in principle one needs to solve an IDE on a triangular time grid $(i\Delt,j\Delt)$ with some small time-step $\Delt$ and $j\leq i$. All functions $f$ exist on this grid, $f_{ij}=f(i\Delt,j\Delt)$ or $f_i=f(i\Delt)$ for one-time functions. In case of the step-force (\ref{eq:force_protocol}) that is considered here one part of the triangle is described by the constant force solution and another one by the equilibrium solution (meaning constant force equal to zero), so what is actually left to be calculated is the square $[t_s,2t_s]\times[0,t_s]$.

In equation (\ref{eq:split_eom}) there are two integrals that need to be written in discrete form. 
For the first one we define $t_k:=t_s-k\Delt$ with $t'=t_j$ and do the following steps
\begin{align}
	&\int_{t'}^{t_s} \! ds \, m(t,s)\partial_s \phi_F(s-t') \nonumber \\
	=~& \sum_{k=j-1}^0 \int_{t_{k+1}}^{t_k}  \! ds \, m(t,s)\partial_s \phi_F(s-t') \nonumber  \\
	\approx~& \sum_{k=0}^{j-1} \frac{m(t,t_k)+m(t,t_{k+1})}{2}  \int_{t_{k+1}}^{t_k}  \! ds \,\partial_s \phi_F(s-t') \nonumber  \\
	=~& \sum_{k=0}^{j-1} \frac{m_{ik}+m_{i,k+1}}{2} \left( \phi^F(t_k-t_j) - \phi^F(t_{k+1}-t_j)\right) \nonumber  \\
	=~& \sum_{k=0}^{j-1} \frac{m_{ik}+m_{i,k+1}}{2} \left( \phi^F_{j-k} - \phi^F_{j-k-1}\right)
\end{align}
For the second one we redefine $t_k:=t_s+k\Delt$ with $t=t_i$ and proceed similarly
\begin{align}
	&\int_{t_s}^t \! ds \, m_\eq (t-s)\partial_s\phi(s,t') \nonumber \\
	=~& \sum_{k=0}^{i-1} \int_{t_k}^{t_{k+1}} \! ds \, m_\eq (t-s)\partial_s\phi(s,t') \nonumber \\
	\approx~& \sum_{k=0}^{i-1} \frac{m^\eq_{i-k}+m^\eq_{i-k-1}}{2} \left( \phi_{k+1,j} - \phi_{k,j}\right)
\end{align}
Lastly we express the derivative as \cite{GruberDiss,FrahsaDiss}
\begin{equation}
    \partial_t\phi(t,t') \approx
    \frac{1}{\Delt}\left(
    \frac{3}{2}\phi_{i,j} - 2\phi_{i-1,j} + \frac{1}{2}\phi_{i-2,j}\right)
\end{equation}
Sorting out the terms this leads to the discretized version of (\ref{eq:split_eom})
\begin{equation}
	A \phi_{ij} = D_{ij} + B m_{ij} + S_{ij}
    \label{eq:2d_scheme}
\end{equation}
with
\begin{gather}
	A_i = \frac{3}{2\Delt}+\frac{m^\eq_{0}+m^\eq_{1}}{2}+1,	\\
	B = \frac12(\phi^F_{0}-\phi^F_{1}), \\
	D_{ij} = \frac{1}{2\Delt}(4\phi_{i-1,j}-\phi_{i-2,j}),	\\
	S_{ij} = \phi_{i-1,j}\frac{m^\eq_{0}+m^\eq_{1}}{2} - \frac{\phi^F_{1}-\phi^F_{0}}{2}m_{i,j-1} \nonumber \\
	- \sum_{k=0}^{j-2}\frac{m_{ik}+m_{i,k+1}}{2} \left( \phi^F_{j-k} - \phi^F_{j-k-1}\right) \\
	- \sum_{k=1}^{i-1}\frac{m^\eq_{i-k}+m^\eq_{i-k+1}}{2} \left( \phi_{k,j} - \phi_{k-1,j}\right). \nonumber
\end{gather}
One needs to note that we have changed the definition of the occurring two-time discrete functions to $\phi_{i,j}=\phi(t_s+i\Delt,t_s-j\Delt)$. We also have used the fact that some of the appearing function can be expressed with one time argument and are in practice pre-calculated in a separate one-time scheme, e.g. \cite{Gruber2016}

The equation above is now solved on the square grid in the following way. Starting at the point $(0,0)$ meaning $(t_s,t_s)$ we progress along the $t'$-axis $(0,j)$ up to the point $(0,N_t)$, where $N_t$ gives the size of the grid, the number of grid points in one direction. Then go through all $(1,j)$ and continue until all points are calculated. The calculation of each point means to make a fixed-point iteration of Eq.~(\ref{eq:2d_scheme}), which requires knowledge of certain previously calculated points, which is ensured by the procedure described above.

To greatly increase the time window that can be calculated by this method we employ \emph{decimation} and repeated execution of the calculation. This means after the square has been fully calculated we define a new grid of $N_t \times N_t$ points only now with the doubled time step $2\Delt$. The first $\frac{N_t}{2}\times\frac{N_t}{2}$ values can be carried over from the previous calculation. The remaining points are calculated in similar order as the ones before. This decimation procedure is repeated $n$ times until $2^n \Delt N_t = t_s$

To obtain the mean displacement from the schematic model we have the additional equation ($t>t_s$)
\begin{align}
	&\partial_t z(t) + \int_0^{t_s} \! ds \, m_z(t,s) \partial_{s} z^F(s) \nonumber \\
	&=- \int_{t_s}^t \! ds \, m_z^\eq(t-s) \partial_{s} z(s),
\end{align}
which can not be solved in every decimation step but only in the last one, when the calculation has reached $t_s$ which is in the integral boundaries. Analogous to the integrals before we use discretizations (here esp. $j=N_t$)
\begin{align}
	&\int_0^{t_s} \! ds \, m_z(t,s) \partial_{s} z^F(s) \nonumber \\
	\approx ~&\sum_{k=0}^{N_t-1} \frac{m^z_{ik}+m^z_{i,k+1}}{2} \left( z^F_{N_t-k} - z^F_{N_t-k-1}\right)
\end{align}
and
\begin{align}
	&\int_{t_s}^t \! ds \, m_z^\eq(t-s) \partial_{s} z(s) \nonumber \\
	&~~\approx \sum_{k=0}^{i-2} \frac{m^{z,\eq}_{i-k}+m^{z,\eq}_{i-k-1}}{2} \left( z_{k+1} - z_{k}\right)\\
	&~~~~+ \frac{m^{z,\eq}_{1}+m^{z,\eq}_{0}}{2} \left( z_{i} - z_{i-1}\right) \nonumber
\end{align}
This again leads to a scheme of the form
\begin{align}
	A z_i = D_i + S_i
\end{align}
with
\begin{gather}
	A = \frac{3}{2\Delt} + \frac{m^{z,\eq}_{1}+m^{z,\eq}_{0}}{2} \\
	D_i = \frac{1}{2\Delt} (4 z_{i-1} - z_{i-2}) \\
	S_i = \frac{m^{z,\eq}_{1}+m^{z,\eq}_{0}}{2} z_{i-1} \nonumber \\
	- \sum_{k=0}^{N_t-1} \frac{m^z_{ik}+m^z_{i,k+1}}{2} \left( z^F_{N_t-k} - z^F_{N_t-k-1}\right)\\
	- \sum_{k=0}^{i-2} \frac{m^{z,\eq}_{i-k}+m^{z,\eq}_{i-k-1}}{2} \left( z_{k+1} - z_{k}\right) \nonumber
\end{gather}
Since the kernel $m_z$ does not need to be determined self-consistently anymore at this point, the solution of the above equation can be calculated without using an iteration.

We want to mention some further details relevant for this implementation of the numerical solution. It occurred useful to make every numerical parameter a power of 2. This is because of the \emph{doubling} of the time step in every decimation window. Thus $\Delt=2^{-a}$, $N_t=2^b$, and also $t_s=2^c$. This means 
\begin{equation}
    2^c = t_s = N_t\cdot 2^n \cdot \Delta_t = 2^{b+n-a} ~ \Rightarrow ~
    n = a+c-b .
\end{equation}
Typical values are e.g. $a=30$, $b=10$ and $c=8$, which would mean we have to make $n=28$ decimation steps. The numerical quality, i.e. the degree of convergence of the result to the actual solution of the IDE is best expressed by the quotient $\frac{N_t}{t_s}=2^{b-c}$ which is the density of points in the last calculation. This means we get the best results if we cessate the force as early as possible e.g. after the system has reached its steady state and increase the overall number of grid points as far as computationally feasible. 

One also needs to note that $t_s$ is in a way determined in retrospect by the number of decimation windows that are calculated. This implies also the restriction that the accessible duration after force shut-off is exactly as long as $t_s$. Although the region of the correlator close to $(t_s,t_s)$ is determined for very small time step, the calculation of $z$ still only can happen on the coarsest grid after all decimation steps, since the integral there starts from zero which decreases numerical precision.

\section{Effect of the solvent friction coefficient in the simulations} \label{app_sims}

\begin{figure}
	\centering
	\includegraphics{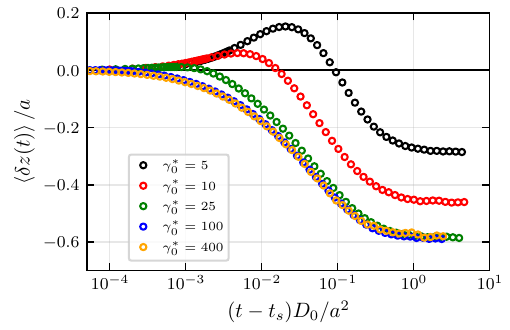}
	\caption{Comparison of the tracer displacement from simulations for different values of $\gamma_0 = \gamma_0^* \sqrt{m k_B T}/a$ after the force shut-off ($F_\ex = 100 \kbT/a$ and $\varphi=0.50$).}
	\label{fig:recoil_sim_gamma0}
\end{figure}

Fig. \ref{fig:recoil_sim_gamma0} studies the tracer displacement after the force is switched off, at time $t_s$, for different values of the friction coefficient with the solvent, $\gamma_0$, for a constant force, $F_\ex=100\,\kbT/a$. The inertia of the tracer is noticed as the increasing trend for short times, until momentum relaxes. This takes longer for smaller $\gamma_0$, as expected, before the elastic response of the bath pushes the tracer back and a constant position is reached finally (in average).

 \begin{figure}
	\centering
	\includegraphics{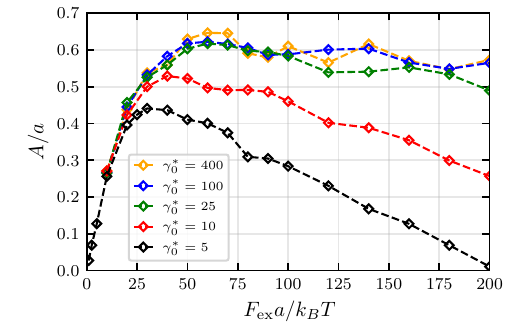}
	\caption{Recoil amplitude as a function of the external force for different values of $\gamma_0$, as labeled.}
	\label{fig:recoil_amp_gamma0}
\end{figure}

The resulting recoil amplitude as a function of the external force is presented in Fig. \ref{fig:recoil_amp_gamma0} for different values of $\gamma_0$. As anticipated previously, the inertial effects at short times result in different amplitudes, until the effect of the Langevin friction $\gamma_0$ saturates (for values above $\approx 50 \sqrt{m\kbT}/a)$. However, note that for small forces, all curves collapse onto a single master curve, where the amplitude grows linearly with the force, i.e. the linear regime. Because the theory considers the overdamped case, one concludes that a solvent friction coefficient of $\gamma_0=100\,\sqrt{m\kbT}/a$ should be used in the following simulations, particularly when large forces are analyzed.

\bibliographystyle{unsrt}
\bibliography{refs}

\end{document}